\documentclass{pasj00}
\begin{document}
\SetRunningHead{INADA ET AL.}{Fifth image of SDSS J1004+4112}
\Received{2004/10/27}
\Accepted{2005/3/14}
\title{Discovery of a Fifth Image of the Large Separation
Gravitationally Lensed Quasar SDSS J1004+4112\footnotemark[*]} 
\author{%
Naohisa \textsc{Inada},\altaffilmark{1}
Masamune \textsc{Oguri},\altaffilmark{2,3}
Charles R. \textsc{Keeton},\altaffilmark{4} 
Daniel J. \textsc{Eisenstein},\altaffilmark{5}\\
Francisco J. \textsc{Castander},\altaffilmark{6}
Kuenley \textsc{Chiu},\altaffilmark{7}
Joseph F. \textsc{Hennawi},\altaffilmark{8}
David E. \textsc{Johnston},\altaffilmark{2}\\
Bartosz \textsc{Pindor},\altaffilmark{9}
Gordon T. \textsc{Richards},\altaffilmark{2}
Hans-Walter \textsc{Rix},\altaffilmark{10}
Donald P. \textsc{Schneider},\altaffilmark{11}
Wei \textsc{Zheng}\altaffilmark{7}
}

\altaffiltext{1}{Institute of Astronomy, Faculty of Science, University
of Tokyo,  
2-21-1 Osawa, Mitaka, Tokyo 181-0015.}
\altaffiltext{2}{Princeton University Observatory, Peyton Hall,
Princeton, NJ 08544, USA.}
\altaffiltext{3}{Department of Physics, University of Tokyo, Hongo 7-3-1,
Bunkyo-ku, Tokyo 113-0033.}
\altaffiltext{4}{Department of Physics and Astronomy, Rutgers
University, 136 Frelinghuysen Road, Piscataway, NJ 08854, USA.} 
\altaffiltext{5}{Steward Observatory, University of Arizona,
933 North Cherry Avenue, Tucson, AZ 85721, USA.}
\altaffiltext{6}{Institut d'Estudis Espacials de Catalunya/CSIC,
Gran Capita 2-4, 08034 Barcelona, Spain.}
\altaffiltext{7}{Department of Physics and Astronomy, 
Johns Hopkins University, \\ 3701 San Martin Drive, Baltimore, MD 21218, USA}
\altaffiltext{8}{Department of Astronomy, University of California at
Berkeley, 601 Campbell Hall, Berkeley, CA 94720, USA.}
\altaffiltext{9}{Department of Astronomy, University of Tronto, 60 St.
George Street, Tronto, Ontario M5S 3H8, Canada.} 
\altaffiltext{10}{Max-Planck Institute for Astronomy, K\"onigstuhl
17, D-69117 Heidelberg, Germany.}
\altaffiltext{11}{Department of Astronomy and Astrophysics, The
Pennsylvania State University, \\ 
525 Davey Laboratory, University Park, PA 16802, USA.} 


\maketitle

\begin{abstract}

We report the discovery of a fifth image in the large separation 
lensed quasar system SDSS~J1004+4112. A faint point source located
0\farcs2 from the center of the brightest galaxy in the lensing cluster
is detected in images taken with the Advanced Camera for Surveys (ACS)
and the Near Infrared Camera and Multi-Object Spectrometer (NICMOS)
on the {\em Hubble Space Telescope}. The flux ratio between the point
source and the brightest lensed component in the ACS image is similar
to that in the NICMOS image. The location
and brightness of the point source are consistent with lens model
predictions for a lensed image. We therefore conclude that the point
source is likely to be a fifth image of the source quasar. In
addition, the NICMOS image reveals the lensed host galaxy of the source
quasar, which can strongly constrain the structure of the lensing
critical curves and thereby the mass distribution of the lensing
cluster.
\end{abstract}

%%%%%%%%%%%%%%%%%%%%%%%%%%%%%%%%%%%%%%%%%%%%%%%%%%%%%%%%%%%
%%%%%%%%%%%%%%%%%%%%%%%%%%%%%%%%%%%%%%%%%%%%%%%%%%%%%%%%%%%
\section{Introduction}
%%%%%%%%%%%%%%%%%%%%%%%%%%%%%%%%%%%%%%%%%%%%%%%%%%%%%%%%%%%
%%%%%%%%%%%%%%%%%%%%%%%%%%%%%%%%%%%%%%%%%%%%%%%%%%%%%%%%%%%

\footnotetext[*]{Based on observations with the NASA/ESA {\em Hubble Space Telescope}, 
obtained at the Space Telescope Science 
Institute, which is operated by the Association of Universities for Research 
in Astronomy, Inc., under NASA contract NAS 5-26555. These observations are 
associated with {\em HST} program 9744. }

Gravitational lensing is a unique tool for exploring the distribution of 
matter, particularly that of dark matter. The recently discovered  
largest separation lensed quasar, SDSS~J1004+4112
\citep{inada03,oguri04a}, has opened a new window for probing dark
matter distributions in the universe.  The quasar was discovered in the
Sloan Digital Sky Survey (SDSS; \cite{york00}; \cite{abazajian04}), and
the lensing hypothesis was confirmed by subsequent observations with the
Subaru 8.2-m telescope and the Keck telescope. The system consists of
four lensed quasar components  ($i'=$18.5, 18.9, 19.4, and 20.1) at
$z=1.734$ and the maximum separation angle between the lensed images is
$14\farcs62$. The lensing object must be a massive object, such as  a
cluster of galaxies, to produce such a large image separation; indeed,
we have identified a $z=0.68$ cluster centered among the four lensed
images. The discovery of a single, cluster-size lensed quasar among the
current SDSS quasars is consistent with the theoretical expectation of
lensing based on the cold dark matter model \citep{oguri04a,oguri04b}.   

SDSS~J1004+4112 is unique in the sense that 1) the quadruple images
place robust constraints on the innermost region of the lensing cluster,
and 2) the lensing cluster is a {\it strong lensing selected}
cluster of galaxies. These features indicate that the mass modeling
of this lens system may offer valuable information on the structure
of clusters of galaxies. The first attempt at modeling the system
with various parametric models revealed the elongated and complicated
mass distribution of the lensing cluster \citep{oguri04a}.
\citet{williams04} recently studied this system using a free-form
reconstruction technique and reached similar conclusions.  However,
important degeneracies between different models remain, and further
follow-up observations are required to determine the mass distribution
more precisely.

Observations using the {\em Hubble Space Telescope} ({\em HST}) can
offer such new data.  In particular, high-resolution {\em HST} images
could be quite effective at detecting lensed images of the quasar
host galaxy, arc or arclet images of other lensed sources, and
perhaps a central or ``odd'' image of the lensed quasar system that
would be expected theoretically (e.g., \cite{burke81}; \cite{rusin02}).
A central image is especially useful for providing tight constraints
on the central mass distribution of the lensing object.  This was indeed
demonstrated by the first discovery of a central image in a lensed
quasar system; \citet{winn04} showed that the central image of
PMN~J1632$-$0033 requires $\beta=1.91\pm0.02$ ($2\sigma$ confidence)
when a power-law density profile $\rho(r)\propto r^{-\beta}$ is
assumed.  Possible central images have also been identified in
lensed arc systems, such as CL 0024+1654 \citep{colley96}, MS
2137.3$-$2353 \citep{gavazzi03}, and A1689 \citep{broadhurst05},
and they also provide important constraints on mass models (see,
e.g., \cite{gavazzi03}).

In this {\em letter}, we present an identification of a fifth
image of the lensed quasar with the Advanced Camera for Surveys 
(ACS; \cite{clampin00}) and the Near Infrared Camera and Multi-Object 
Spectrometer (NICMOS; \cite{thompson92}) installed on the {\em HST}.  
In addition, we report unambiguous detection of the lensed host galaxy
in the NICMOS image.

%%%%%%%%%%%%%%%%%%%%%%%%%%%%%%%%%%%%%%%%%%%%%%%%%%%%%%%%%%%
%%%%%%%%%%%%%%%%%%%%%%%%%%%%%%%%%%%%%%%%%%%%%%%%%%%%%%%%%%%
\section{The ACS Observation}
%%%%%%%%%%%%%%%%%%%%%%%%%%%%%%%%%%%%%%%%%%%%%%%%%%%%%%%%%%%
%%%%%%%%%%%%%%%%%%%%%%%%%%%%%%%%%%%%%%%%%%%%%%%%%%%%%%%%%%%

An ACS observation (with the Wide Field Channel) in the F814W filter 
(${\approx}I$-band) was conducted on 2004 April 28, under the program
``{\em HST} Imaging of Gravitational Lenses'' (GO-9744, PI C.\ Kochanek).
The observation
consisted of five dithered exposures taken in ACCUM mode. The total
exposure time was 405 seconds. The reduced (drizzled and calibrated)
images were extracted using the CALACS pipeline \citep{hack1999}, which 
includes the PyDrizzle algorithm. We further rejected cosmic rays using 
the L.A.Cosmic package \citep{dokkum01} in the drizzled image. A
$40''\times40''$ subsection of a median of the five dithered images is
shown in Figure \ref{fig:acs}. Following Figure~9 of \citet{oguri04a},
the four lensed components are denoted as ``A--D'', and three central
bright galaxies of the lensing cluster are denoted as ``G1--G3''. 
The redshifts of
galaxies G1--G3 are 0.680, 0.675, and 0.675, respectively
\citep{oguri04a}. The relative positions of components A--D were
calculated by a single Gaussian fit, and the flux ratios of components
A--D were estimated by fitting the PSF stars produced by the Tiny Tim
software (version 6.1a; \cite{krist03}), in the drizzled (and cosmic
ray rejected) image.  The PSF of a quasar was constructed with an
$\alpha_{\nu}=-0.5$ power law spectrum in the F814W wavelength region
(corresponding to $\sim$3000 {\AA} in the rest frame). The relative
positions derived from the ACS (F814W) are consistent with those
derived from the Subaru $i'$ band  image \citep{oguri04a} within
${\sim}5{\sigma}$. The four components are all unsaturated, and the AB 
magnitude of component A in the F814W filter is estimated to be 18.4.
The position of component G1 was extracted by the Source Extractor
algorithm \citep{bertin96}. The results are summarized in Table
\ref{table:posflux}.  

\begin{figure*}[t]
   \begin{center}
      \FigureFile(100mm,90mm){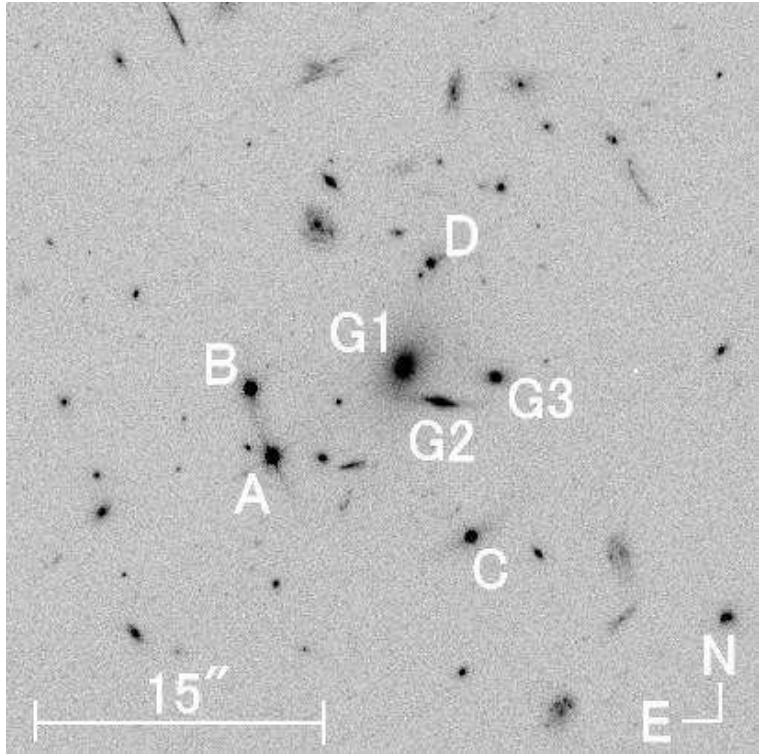}
   \end{center}
   \caption{$40''\times40''$ subsection of the median combined ACS image
    of SDSS J1004+4112. The four quasar images are labeled A--D. 
    The galaxies of the lensing cluster labeled G1--G3 have redshifts of 
    0.680, 0.675, and 0.675, respectively. The pixel scale is
    approximately 0\farcs05 ${\rm pixel^{-1}}$.\label{fig:acs}}
\end{figure*}

\begin{table}
  \caption{Relative Positions and Flux Ratios of SDSS~J11004+4112 in the
 {\em HST} ACS Image\label{table:posflux}}
  \begin{center}
    \begin{tabular}{crrc}
     \hline\hline
      Object & $x$[arcsec]\footnotemark[$*$] &
     $y$[arcsec]\footnotemark[$*$] & Flux Ratio\footnotemark[$\dagger$]\\
     \hline
     A &   0.0000$\pm$0.001  &  0.0000$\pm$0.001 & 1.000 \\
     B &  -1.317$\pm$0.002    &  3.532$\pm$0.002   & 0.732 \\
     C &  11.039$\pm$0.002    & -4.492$\pm$0.002  & 0.346 \\
     D &   8.399$\pm$0.004    &  9.707$\pm$0.004  & 0.207 \\
     E &   7.197$\pm$0.009    &  4.603$\pm$0.009  & 0.003 \\
     G1 &  7.114$\pm$0.030    &  4.409$\pm$0.030  & --- \\
    \hline
     \multicolumn{4}{@{}l@{}}{\hbox to 0pt{\parbox{90mm}{\footnotesize
       \footnotemark[$*$] The positive directions of $x$ and $y$ are
     defined by West and North, respectively.
       \par\noindent
       \footnotemark[$\dagger$] Errors of fitting by quasar PSFs are
     about 10\%. The error of component E might be much larger due to
     the over-subtraction of G1. }\hss}}
\end{tabular}
  \end{center}
\end{table}

%%%%%%%%%%%%%%%%%%%%%%%%%%%%%%%%%%%%%%%%%%%%%%%%%%%%%%%%%%%
%%%%%%%%%%%%%%%%%%%%%%%%%%%%%%%%%%%%%%%%%%%%%%%%%%%%%%%%%%%
\section{The NICMOS Observation}
%%%%%%%%%%%%%%%%%%%%%%%%%%%%%%%%%%%%%%%%%%%%%%%%%%%%%%%%%%%
%%%%%%%%%%%%%%%%%%%%%%%%%%%%%%%%%%%%%%%%%%%%%%%%%%%%%%%%%%%

\begin{figure}[t]
   \begin{center}
      \FigureFile(80mm,90mm){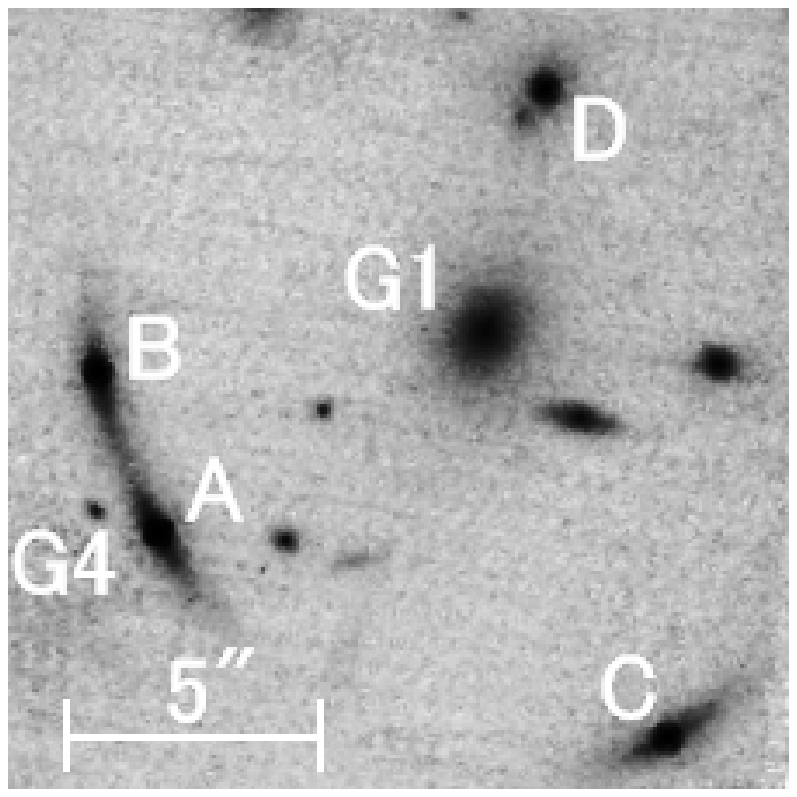}
   \end{center}
   \caption{The combined NICMOS image of SDSS J1004+4112. 
    The pixel scale is approximately 0\farcs075 ${\rm pixel^{-1}}$. 
    The lensed host galaxy is seen more prominently in this NICMOS 
    image than in the ACS image. In addition, we can see G4 
    near component A. 
    \label{fig:nicmos}}
\end{figure}

The NICMOS imaging observation was also conducted under the same
{\em HST} program on 2004 October 9. The
observation consists of four dithered exposures taken in MULTIACCUM mode,
using the F160W filter (${\approx}H$-band). The exposure time was 640 
sec for two of the exposures and 704 sec for the other two. 
The calibration was extracted by 
the CALNICA pipeline, and the central bad columns of each dithered image 
were corrected by linear interpolation. The combined image is shown 
in Figure \ref{fig:nicmos}. First, we confirm a galaxy near 
component A (marked as G4 in Figure \ref{fig:nicmos}), which may host 
the star or stars responsible for microlensing of the broad emission 
line region \citep{richards04}. In addition, extended emission is
clearly seen around components A, B and C. Although such extensions are 
also seen in the ACS images (see Figure \ref{fig:acs}), their existence 
is much more robust in the NICMOS image. The fact that these extensions
are obvious in the F160W (near-infrared) image and faint in the F814W
(optical) image, and that the distortions agree with the theoretical
critical curves (see Figure 17 of \cite{oguri04a}), demonstrates that
the extended flux is due to the lensed host galaxy of the source quasar. 
These images provide many new constraints on lens models that will
significantly improve our ability to determine the mass distribution of
the lensing cluster. They do make lens modeling more computationally
intensive (because one must account for the intrinsic shape of the host
galaxy, and also for the effects of the point spread function), so we
defer detailed modeling of the arcs to a subsequent paper.

%%%%%%%%%%%%%%%%%%%%%%%%%%%%%%%%%%%%%%%%%%%%%%%%%%%%%%%%%%%
%%%%%%%%%%%%%%%%%%%%%%%%%%%%%%%%%%%%%%%%%%%%%%%%%%%%%%%%%%%
\section{Fifth Image}
%%%%%%%%%%%%%%%%%%%%%%%%%%%%%%%%%%%%%%%%%%%%%%%%%%%%%%%%%%%
%%%%%%%%%%%%%%%%%%%%%%%%%%%%%%%%%%%%%%%%%%%%%%%%%%%%%%%%%%%

\begin{figure}[t]
   \begin{center}
    \FigureFile(40mm,*){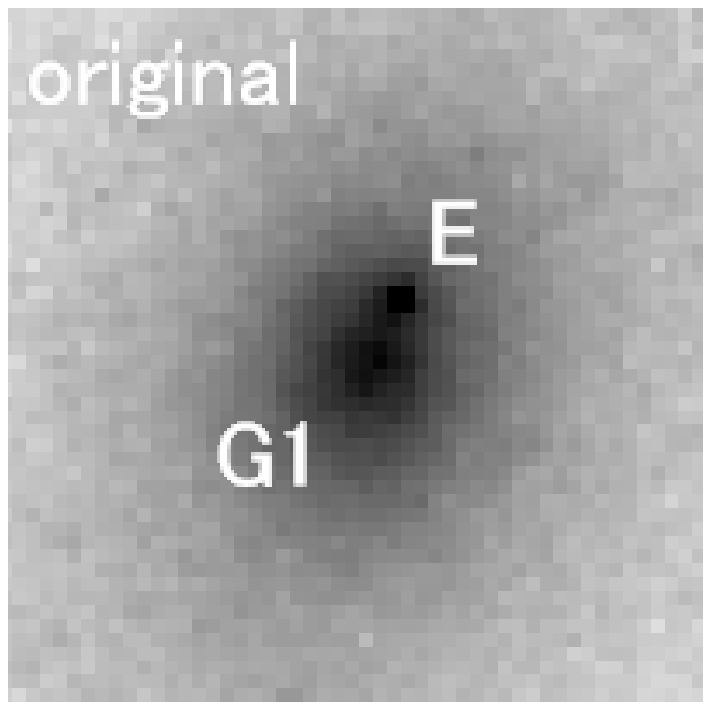}
    \FigureFile(40mm,*){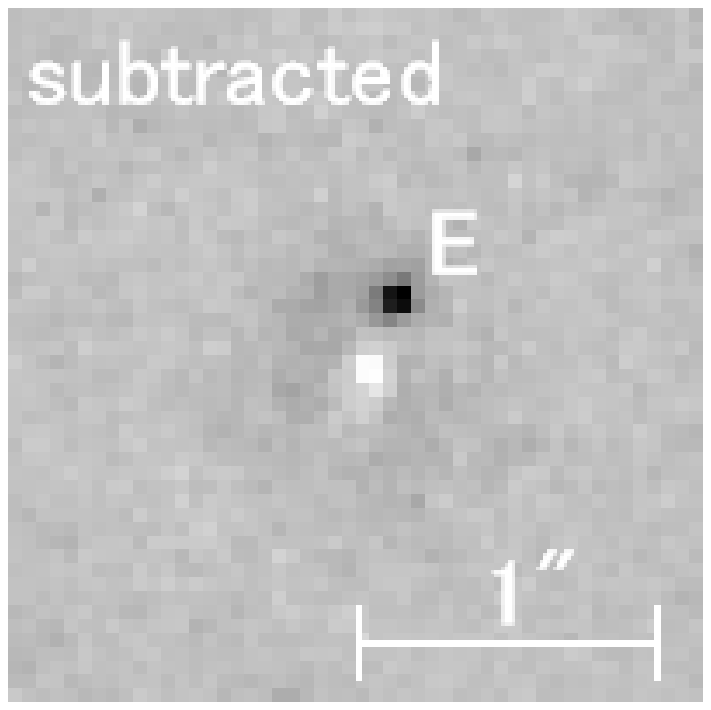}
   \end{center}
   \caption{Left: The ACS image of central galaxy G1. Right: The ACS
 image after subtracting galaxy G1. The residual image clearly shows
 a stellar object near the center of G1.\label{fig:acsfifth}}
\end{figure}

Of interest is the existence of a point source near the center of G1.
The left panel in Figure \ref{fig:acsfifth} shows the ACS image of 
galaxy G1; what appears to be an unresolved source is clearly seen
approximately 0\farcs2 northwest of the center of G1. {\em This feature is
neither a bad pixel nor a cosmic ray}; the source is seen in all
dithered images. The right panel in Figure \ref{fig:acsfifth} displays the
image after subtracting the signal from G1 
(modeled with the GALFIT package of \cite{peng02}).
Due to the existence of the unresolved
source near the center of G1 (the peak flux of this unresolved source
is almost same as that of G1), G1 was slightly over-subtracted.
However, we can see a new source, labeled E, in the subtracted image.
This object is classified as a point source by the Source Extractor
algorithm. The position and brightness of E, based on a single Gaussian
fit to the data, are given in Table \ref{table:posflux}.  

\begin{figure}[t]
   \begin{center}
    \FigureFile(40mm,*){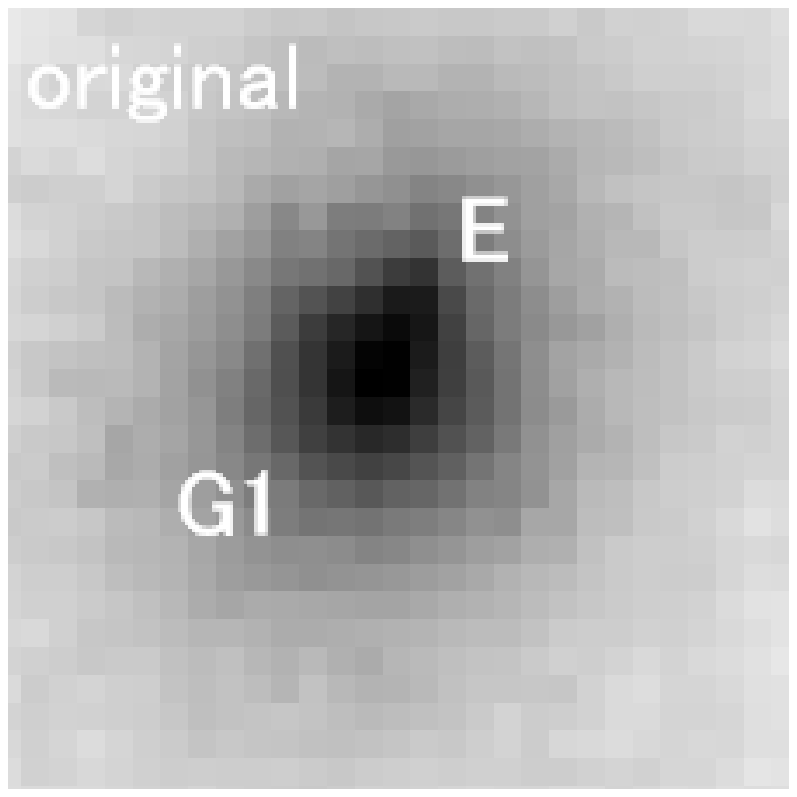}
    \FigureFile(40mm,*){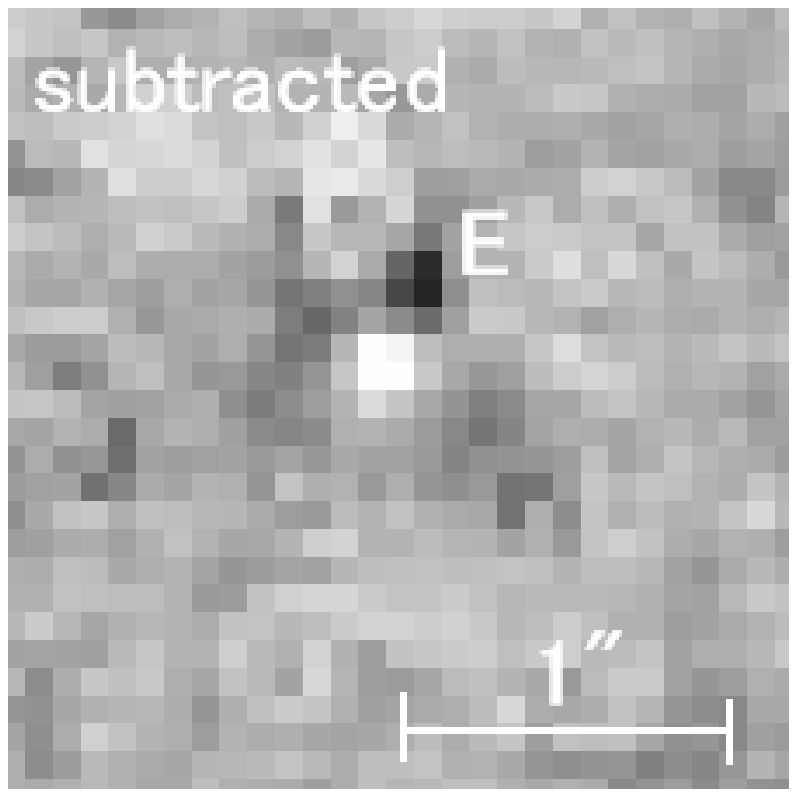}
   \end{center}
   \caption{Left: The NICMOS image of central galaxy G1. Right: The NICMOS
 image after subtracting galaxy G1. The stellar object near the center of 
 G1 can be also seen in the NICMOS image. \label{fig:nicfifth}}
\end{figure}

We find component E also in the NICMOS image; the left panel in Figure
\ref{fig:nicfifth} is the NICMOS image of galaxy G1. We 
subtracted the signal from G1 using the GALFIT package, which is shown
in the right panel of Figure \ref{fig:nicfifth}. We confirm
component E in the subtracted image, although G1 was slightly 
over-subtracted as in the ACS image. 
Measuring the flux of component E with a single Gaussian fit, we find
that the flux ratios between E and A ($E/A$) in the ACS and NICMOS images 
are 0.003 and 0.004, respectively. This remarkable agreement of the flux 
ratios supports the idea that the point source is a fifth image of the 
lensed quasar. 

\begin{figure}[t]
   \begin{center}
    \FigureFile(80mm,90mm){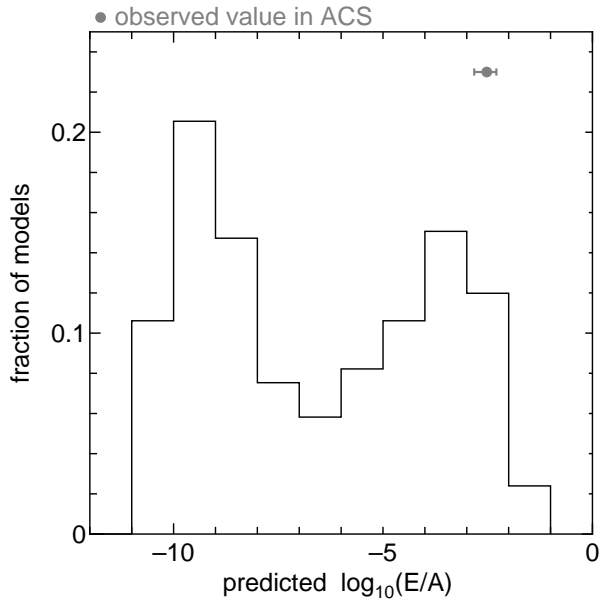}
   \end{center}
   \caption{
 Frequency (fraction of models) of the predicted flux ratio of $E/A$ from 
 two component (central galaxy + cluster component) lens models.  
 The models are constrained by the observed positions and brightnesses
 of components A--D (and the observed position of G1), 
 and the position of component E. We assume that
 the scale radius of the NFW component is 40\farcs0, but the results
 are insensitive to the particular value.  The gray filled circle
 represents the observed value of $E/A$ (in ACS) with 50\% error. 
\label{fig:model}}
\end{figure}

To test this hypothesis, we have refined the lens models presented in
\citet{oguri04a} using the more precise {\em HST} data. The models
consist of a singular isothermal ellipsoid mass distribution for
galaxy G1, and an NFW \citep{navarro97} elliptical potential for the
cluster.  We demand that the models reproduce the relative positions
and brightnesses of components A--D (and the relative position of G1),
as well as the position of
component E; we do not use the brightness of E as a constraint,
because we want to see what the models predict. Adopting the same approach
as \citet{oguri04a}, we use the {\em lensmodel} software
\citep{keeton01} for Monte Carlo sampling of the parameter space.
There is a wide range of models consistent with the data, indicating
that it is not difficult to produce a 5-image lens matching the
configuration of components A--E, and that significant model
degeneracies remain.  More interesting are the model predictions
for the flux ratio between the fifth image and component A, shown
in Figure \ref{fig:model}.  The predictions span a remarkable nine
orders of magnitude, from $E/A \sim 0.1$ down to $E/A \sim 10^{-10}$,
but a significant fraction predict $E/A$ in the range 0.001--0.01,
consistent with the observed value.  In other words, there are
(many) reasonable models that can fit all of the {\em HST} data
under the hypothesis that E is a fifth image; conversely, the
observed properties of E are highly compatible with that hypothesis.

Given the enormous range of model predictions for the brightness
of E, it appears that the observed brightness offers strong
constraints on the models.  We caution that one must be careful
in using the brightness of E as a constraint, because its measured
flux could be contaminated by improper subtraction of the galaxy
or by physical effects such as microlensing or extinction.
Nevertheless, the range of predictions is so large that even
conservative estimates of systematic uncertainties should still
yield very interesting results.  As an example, to test the ability
of E to constrain the central density profile of G1, we switch from
an isothermal model to a more general power law density profile
$\rho(r)\propto r^{-\beta}$ for the galaxy (for computational
simplicity, we now assume that the {\em potential}, rather than
the density, has elliptical symmetry).  We find that there is a
wide range of models with $1.6 \le \beta \le 2.0$ that predict
$0.001 \lesssim E/A \lesssim 0.01$.  By contrast, all of the models
we examined with $\beta \ge 2.1$ predict $E/A \lesssim 10^{-10}$,
which grossly contradicts even a conservative reading of the data.
In other words, the observed properties of E imply that the galaxy
mass distribution cannot be steeper than isothermal (i.e.,
$\beta \le 2$).  This upper bound is similar to that found by
\citet{winn04} from the central image in PMN J1632$-$0033
(specifically, they found $\beta = 1.91\pm0.02$).  However,
SDSS J1004+4112 differs from PMN J1632$-$0033 in that we do not
obtain any {\em lower} bound on $\beta$, at least over the range
$1.6 \le \beta \le 2.0$ that we have explored so far.  Apparently
the complexity of the SDSS J1004+4112 lens potential, with a cluster
in addition to the galaxy, prevents a unique measurement of the
value of $\beta$.  Nevertheless, it is clear that component E
provides important new constraints on the mass distribution of
this interesting lens system.

%%%%%%%%%%%%%%%%%%%%%%%%%%%%%%%%%%%%%%%%%%%%%%%%%%%%%%%%%%%
%%%%%%%%%%%%%%%%%%%%%%%%%%%%%%%%%%%%%%%%%%%%%%%%%%%%%%%%%%%
\section{Summary\label{sec:sum}}
%%%%%%%%%%%%%%%%%%%%%%%%%%%%%%%%%%%%%%%%%%%%%%%%%%%%%%%%%%%
%%%%%%%%%%%%%%%%%%%%%%%%%%%%%%%%%%%%%%%%%%%%%%%%%%%%%%%%%%%

We have presented the {\em HST} ACS and NICMOS images of SDSS~J1004+4112,
which reveal a fifth image of the lensed quasar core.
The fifth image 
offers a unique probe of the mass distribution of the cluster core.
Deep spectroscopic observations of component E with large telescopes,
such as the Subaru Telescope, offer the best prospect for the final
confirmation that it is a lensed quasar image.  In the NICMOS image, 
we also found unambiguous evidence of the lensed host galaxy of the 
source quasar.  These extended images provide strong additional 
constraints on mass models of the lensing cluster, which are expected 
to break degeneracies seen in the modeling studies to date. A detailed
analysis of lens models including the host galaxy images is
underway and will be presented elsewhere.

%%%%%%%%%%%%%%%%%%%%%%%%%%%%%%%%%%%%%%%%%%%%%%%%%%%%%%%%%%%%%%%
%%%%%%%%%%%%%%%%%%%%%%%%%%%%%%%%%%%%%%%%%%%%%%%%%%%%%%%%%%%%%%%
\bigskip
%%%%%%%%%%%%%%%%%%%%%%%%%%%%%%%%%%%%%%%%%%%%%%%%%%%%%%%%%%%%%%%
%%%%%%%%%%%%%%%%%%%%%%%%%%%%%%%%%%%%%%%%%%%%%%%%%%%%%%%%%%%%%%%

A portion of this work was supported by NASA HST-GO-09744.20. NI and MO
are supported by JSPS through JSPS Research  Fellowship for Young
Scientists. 

Funding for the creation and distribution of the SDSS Archive has been 
provided by the Alfred P. Sloan Foundation, the Participating Institutions, 
the National Aeronautics and Space Administration, the National Science 
Foundation, the U.S. Department of Energy, the Japanese Monbukagakusho, 
and the Max Planck Society. The SDSS Web site is http://www.sdss.org/. 

The SDSS is managed by the Astrophysical Research Consortium (ARC) 
for the Participating Institutions. The Participating Institutions 
re The University of Chicago, Fermilab, the Institute for Advanced 
Study, the Japan Participation Group, The Johns Hopkins University, 
the Korean Scientist Group, Los Alamos National Laboratory, the 
Max-Planck-Institute for Astronomy (MPIA), the Max-Planck-Institute for 
Astrophysics (MPA), New Mexico State University, University of 
Pittsburgh, University of Portsmouth, Princeton University, the 
United States Naval Observatory, and the University of Washington.

%%%%%%%%%%%%%%%%%%%%%%%%%%%%%%%%%%%%%%%%%%%%%%%%%%%%%%%%%%%%%%%%%%%%%%%
%\clearpage
%%%%%%%%%%%%%%%%%%%%%%%%%%%%%%%%%%%%%%%%%%%%%%%%%%%%%%%%%%%%%%%%%%%%%%%

%%%%%%%%%%%%%%%%%%%%%%%%%%%%%%%%%%%%%%%


\begin{thebibliography}{}
%\parskip=-1pt
%\baselineskip=14pt

\bibitem[Abazajian et al.(2004)]{abazajian04}
Abazajian, K. 2004, \aj, 128, 502

\bibitem[Bertin \& Arnouts(1996)]{bertin96}
Bertin, E., \& Arnouts, S. 1996, A\&AS, 117, 393

\bibitem[Broadhurst et al.(2005)]{broadhurst05}
Broadhurst, T., et al.\ 2005, \apj, in press (astro-ph/0409132) 

\bibitem[Burke(1981)]{burke81}
Burke, W.~L.\ 1981, \apjl, 244, L1 

\bibitem[Clampin(2000)]{clampin00}
Clampin, M. et al. 2000, SPIE, 401, 344

\bibitem[Colley, Tyson, \& Turner(1996)]{colley96} 
Colley, W.~N., Tyson, J.~A., \& Turner, E.~L.\ 1996, \apjl, 461, L83 

\bibitem[van Dokkum(2001)]{dokkum01}
van Dokkum P. G. 2001, \pasp, 113, 1420

\bibitem[Gavazzi et al.(2003)]{gavazzi03}
Gavazzi, R., Fort, B., Mellier, Y., Pello, R., \& Dantel-Fort, M.
2003, \aap, 403, 11

\bibitem[Hack(1999)]{hack1999}
Hack, W. J. 1999, CALACS Operation and Implementation, ISR ACS-99-03

\bibitem[Inada et al.(2003)]{inada03}
Inada, N., et al.\ 2003, \nat, 426, 810

\bibitem[Keeton(2001)]{keeton01}
Keeton, C.~R.\ 2001, preprint (astro-ph/0102340)

\bibitem[Krist \& Hook(2003)]{krist03}
Krist, J. E. \& Hook, R. N. 2003, The Tiny Tim User's Guide, Version 6.1a

\bibitem[Navarro, Frenk, \& White(1997)]{navarro97}
Navarro, J. F., Frenk, C. S., \& White, S. D. M. 1997, \apj, 490, 493

\bibitem[Oguri et al.(2004)]{oguri04a}
Oguri, M., et al. 2004, \apj, 605, 78

\bibitem[Oguri \& Keeton(2004)]{oguri04b}
Oguri, M., \& Keeton, C.~R. 2004, \apj, 610, 663

\bibitem[Peng et al.(2002)]{peng02}
Peng, C. Y., Ho, L. C., Impey, C. D., \& Rix, H.-W. 2002, \aj, 124, 266

\bibitem[Richards et al.(2004)]{richards04}
Richards, G.~T., et al. 2004, \apj, 610, 679

\bibitem[Rusin(2002)]{rusin02}
Rusin, D. 2002, \apj, 572, 705

\bibitem[Thompson(1992)]{thompson92}
Thompson, R. 1992, Space Science Reviews (ISSN 0038-6308), 61, no. 1-2, 69

\bibitem[Williams \& Saha(2004)]{williams04}
Williams, L.~L.~R.~\& Saha, P.\ 2004, \aj, 128, 2631

\bibitem[Winn, Rusin, \& Kochanek(2004)]{winn04}
Winn, J.~ N., Rusin, D., \& Kochanek, C.~ S. 2004, \nat, 427, 613

\bibitem[York et al.(2000)]{york00}
York, D.~G., et al. 2000, \aj, 120, 1579

\end{thebibliography}
\end{document}